%% file: paperMain.tex
\documentclass[%
 reprint,
nofootinbib,
 amsmath,amssymb,
 aps,
 pre,
floatfix,
]{revtex4-2}
\usepackage{graphicx}
\usepackage{subcaption}
\usepackage[title]{appendix}
\usepackage[english]{babel}
\usepackage{hyperref}
\input{orcidSetup.tex}
\graphicspath{ {graphs/} }

\begin{document}

\title{Turbulence and Capillary Waves on Bubbles}
\author{Peleg Emanuel\orcidPeleg}
\email{peleg.emanuel@mail.huji.ac.il}
\author{Alexander Feigel\orcidSasha}
\email{alexander.feigel@mail.huji.ac.il}
\affiliation{Racah Institute of Physics, The Hebrew University, 9190401 Jerusalem, Israel}

\input{paperAbstract.tex}

\maketitle

\input{paperIntro.tex}
\input{paperResults.tex}
\input{paperDiscussion.tex}

\bibliographystyle{unsrt}
\bibliography{references}

\pagebreak
\onecolumngrid
\appendix\newpage\markboth{Appendix}{Appendix}
\numberwithin{equation}{section}

\begin{appendices}

\input{paperAppendixGaunt.tex}
\newpage
\input{paperAppendixGeometry.tex}

\newpage
\input{paperAppendixDerivation.tex}
\newpage
\input{paperAppendixNumericalSchemes.tex}
\newpage
\input{paperAppendixTestProblems.tex}
\newpage
\input{paperAppendixInitialConditions.tex}
\newpage
\input{paperAppendixDispAndPump.tex}

\end{appendices}

\end{document}

%% file: orcidSetup.tex
\usepackage{xcolor}
\usepackage{fontawesome5}

\definecolor{orcidLogoColour}{HTML}{A6CE39}

\newcommand{\orcid}[1]{\href{https://orcid.org/#1}{{\color{orcidLogoColour}\faOrcid{}}}}
\newcommand{\orcidPeleg}[0]{\orcid{0000-0002-4503-6715}}
\newcommand{\orcidSasha}[0]{\orcid{0000-0002-5394-4668}}

%% file: paperAbstract.tex
\begin{abstract}
    We present a link between the theory of deep water waves
    and that of bubble surface perturbations. 
    Theory correspondence is shown analytically for small wavelengths
    in the linear regime 
    and investigated numerically in the nonlinear regime. 
    To do so, we develop the second-order spatial perturbation equations 
    for the Rayleigh-Plesset equation and solve them numerically.
    Our code is publicly available. 
    Studying capillary waves on stable bubbles, 
    we recreate the Kolmogorov-Zakharov spectrum 
    predicted by weak turbulence theory, 
    putting wave turbulence theory to use for bubbles. 
    In this investigation, 
    it seems that curvature does not affect turbulent properties. 
    Calculated bubble surface qualitatively responds 
    to low gravity experiments. 
    The link demonstrated opens new possibilities 
    for studying several bubble phenomena, 
    including sonoluminescence and cavitation, 
    using the extensive tools developed in the wave turbulence framework.
\end{abstract}

%% file: paperIntro.tex
\section{Introduction}

The Rayleigh-Plesset equation, first developed by Lord Rayleigh \cite{frs_viii_1917}
governs the dynamics of spherically symmetric bubbles.
It is often used in the study of cavitation \cite{plesset_bubble_1977}
and sonoluminescense \cite{brenner_acoustic_1996}.
First-order stability analysis has been carried out by Plesset as early as the fifties
\cite{plesset_stability_1954}. However, since it predicts the existence
of instabilities, one has to include higher-order terms.
This necessity arises in numerous hydrodynamic systems.
However, investigations of nonlinear instability growth are conducted
primarily numerically on limited systems
\cite{wang_weakly_2015, zhang_weakly_2018, zhao_two-dimensional_2019}.
In this work, we connect the theory of surface hydrodynamic instabilities
to that of wave turbulence in the hope of providing
additional theoretical support to such investigations.

The theory of wave turbulence has been subject to many studies.
It predicts the emergence
of several turbulent spectra, including the well known
Kolmogorov-Zakharov spectra
\cite{zakharov_kolmogorov_2012},
which have been verified numerically and experimentally
\cite{deike_direct_2014, pushkarev_turbulence_1996, falcon_capillary_2009, berhanu_space-time-resolved_2013, dyachenko_weak_2004}.
In certain limiting cases, it is possible to replace
equations with effective Nonlinear Schrödinger Equations,
and analyze stability analytically
\cite{zakharov_stability_1972}.
Many of these results are universal, allowing analogies
between different branches of physics.

Deepwater waves have been studied
in wave turbulence framework, primarily by Zakharov
\cite{zakharov_weak_1971,zakharov_stability_1972,zakharov_kolmogorov_2012}.
The fundamental forces governing their dynamics are surface tension 
and gravity.
Recalling the gravity-acceleration duality,
these are also the forces governing
the dynamics of other fluid interfaces, bubbles included.
Bubbles, however, are curved, and their surface
acceleration may vary both spatially and temporarily
unlike the earth's constant gravity acceleration.

A proof of this analogy in the linear regime is easy to obtain.
In the classical paper\cite{plesset_stability_1954}, Plesset develops
the linear perturbation equations of motion
for an almost spherical surface between two incompressible fluids,

\begin{equation}
    \ddot a_n 
    + 3 \frac{\dot R}{R} \dot a_n
    + W_n a_n
    = 0
    .
\end{equation}

Here $R$ is the mean radius, and $a_n$ denotes the amplitude of some mode with angular momentum number $n$.
Denote by $\sigma$ the surface tension coefficient
and by $\rho_{in / out}$ 
the densities inside and outside the bubble, respectively.
$W_n$ takes the form

\begin{equation}
    W_n = \frac{O\left( \frac{n}{R}\right) 
    \left( - \ddot R \right) \left(\rho_{out} - \rho_{in}\right)
    + O\left(\left(\frac{n}{R}\right)^3\right) 
    \sigma}{\rho_{in} + \rho_{out}}
    .
\end{equation}

For $n > 0$, $a_n$ dynamics are much
more rapid than those of the radius,
which can be regarded as quasistatic.
Thus, for $W_n > 0$ the equation discribes a damped / pumped harmonic oscillator.
Choice of damping or pumping is determined by the sign of $ \dot R$.
For empty bubbles, plugging in $\rho_{in} = 0$
we find that the oscillation frequency, 
$\sqrt{W_n}$, follows the well known
gravity-capillary spectrum $\omega_k = \sqrt{gk + \alpha k^3}$, 
Here $k$ is the wavenumber, $g$ gravity's constant acceleration, 
and $\alpha \equiv \sigma / \rho_l$ the capillary coefficient.
For bubbles we simply plug in $g = -\ddot R$,
which is the fictitious acceleration exerted on an observer moving with the radius,
and $k \sim n / R$.

If acceleration and density gradient are of different signs,
and surface tension is small enough,
it might be that $W_n < 0$. 
In such a case, perturbation growth is exponential
with the rate $\sqrt{gk}$,
corresponding to the linear growth rate of Rayleigh-Taylor instabilities.

We thus conclude that bubble spatial perturbations of small wavelength
follow the behavior of either 
Rayleigh-Taylor instabilities or deep water waves,
at least in first-order.
In this work, we investigate this correspondence in the nonlinear regime.
The most convenient demonstration of this link is the study of capillary waves on a stable bubble.
Weak turbulence theory predicts that $4$-wave interactions dominate gravity wave turbulence \cite{zakharov_kolmogorov_2012}.
Therefore, gravity turbulence simulation
should require third-order perturbation equations of motion,
derived from a Hamiltonian of order $4$.
On the other hand, capillary wave turbulence is 
$3$-wave interaction dominant \cite{zakharov_kolmogorov_2012}. 
Hence, a second-order analysis should suffice for the latter, 
making it a preferred candidate.

We make a second-order analysis of the Rayleigh-Plesset equation
for empty bubbles.
Simulating the resulting equations numerically
we expect to find the theoretical capillary
Kolmogorov-Zakharov elevation
spectrum\footnote{
    Note that sometimes one equivalently refers to the spectrum of wave density,
    $n_k ~ \sim k^{ - \frac{17}{4}}$.},
given by
\cite{zakharov_kolmogorov_2012,zakharov_weak_1971}

\begin{equation}
    <|\eta_k|^2 > \sim k^{ - \frac{19}{4}} \sim \omega_k^{ - \frac{19}{6}}
    .
\end{equation} 

This result has been verified both numerically and experimentally
\cite{deike_direct_2014, pushkarev_turbulence_1996, 
berhanu_space-time-resolved_2013, falcon_capillary_2009},
but not for bubbles.
In \cite{falcon_capillary_2009}, 
low gravity experiments 
were conducted on a system similar to ours.
Capillary waves were observed on the spherical water surface,
and a Kolmogorov-Zakharov power-law matching
the prediction by weak turbulence theory was measured.
This finding suggests curvature has little to no effect on the spectrum.
We set out to corroborate these results numerically.

Having established a connection, 
we hope the extensive theories developed 
in wave turbulence framework 
will be of aid in the study of bubble phenomena.

%% file: paperResults.tex
\section{Results}

We derive the bubble surface equations of motion from hydrodynamic principles, 
following the footsteps of Plesset \cite{plesset_stability_1954}. 
We address bubbles that do not "fold", i.e., bubbles 
whose surface may be defined by $r = \psi\left(\Omega\right)$. 
Here $r$ and $\Omega$ are the radius and solid angle, respectively.
We call $\psi$ the local radius and the shape it defines a semisphere. 
We may expand the local radius in spherical harmonics

\begin{gather}
    \psi\left(\Omega\right) = R + \sum_{l, m} Y_{lm} \left(\Omega\right) a_{lm}
    .
\end{gather}

Here $Y_{lm}$ denote spherical harmonics, 
$a_{lm}$ perturbation amplitudes and $R$ the mean radius. 
When summing over $\left(l,m\right)$ 
we implicity sum all $-l \le m \le l$ and $l > 0$.

For simplicity, we assume
the exterior is far denser than the interior.
As a result, terms containing the density and velocities of the inner fluid are negligible, much like the air above water waves. 
We assume the exterior consists of an incompressible fluid,
and that flow is potential.
Therefore, the hydrodynamic potential, 
denoted $\phi$, obeys Laplace's equation
and can similarly be expanded outside the bubble

\begin{gather}
    \phi\left(r,\Omega\right) = \frac{\dot R R^2}{r} 
            + \sum_{lm} \frac{R}{l + 1}Y_{lm} \left(\Omega\right) b_{lm} 
            \left(\frac{r}{R}\right)^{ -\left(l + 1\right)}
            .
\end{gather}

The bubble's surface is the
interface separating two materials.
Hence, the mass flux through the bubble is 
zero\footnote{
  For non-empty bubbles, this should hold,
  separately, for masses of both materials.},
namely

\begin{gather}
    \label{eq:volumetic flux}
    \vec n \left(\Omega\right) \cdot \left(\dot \psi \hat r
            - \vec v \left(\psi\left(\Omega\right), \Omega)\right)\right) = 0
            .
\end{gather}

Here $\vec v$ denotes the fluid velocity and $\vec n$ the normal to the surface.
Much like Plesset \cite{plesset_stability_1954},
we now evaluate the pressure on both sides using the Bernoulli
integral and by setting the force exerted on the zero mass surface to zero.
Although we assume that the interior density is small, 
we allow inner pressure.
Denote by ${P}_{in/out}$ the inner and outer pressures, respectively,
and by $\rho_l$ the fluid density.
We find

\begin{gather}
  \label{eq: eom base}
    P_{out} + \rho_l \left(\frac{\partial \phi}{\partial t} - \frac{1}{2} v^2\right) |_{r =\psi} = 
    P_{in} - \sigma \left(\frac{1}{R_1} + \frac{1}{R_2}\right)
    .
\end{gather}

Here $\sigma$ is the surface tension coefficient and $R_{1}, R_{2}$ the principle radii.
We define the spherically symetric pressure term 
$\Xi = \frac{1}{\rho_l} \left(P_{out} - P_{in}\right)$,
and use the previously defined capillary coefficient $\alpha = \sigma / \rho_{l}$.
We assume the outer pressure to be a constant $P_\infty$, and the interior
an ideal gas, such that $P_{in} = P_0 \cdot V_0 / V\left(\psi\right)$. 
$P_0$ and $V_0$ acting as reference pressure and volume and 
$V\left(\psi\right)$ the actual semisphere volume.
Our derivation is independent of this assumption and
treats $\Xi$ as a black box. One may plug in any other equation of state.
Expanding equation \eqref{eq: eom base}
in second-order, we find the following equations of motion

\begin{gather}
  \label{eq: perturbation eom}
  \begin{aligned}
      \ddot a_{l,m} & + 3 \frac{\dot R}{R} \dot a_{l,m} - A_{l} a_{l,m} = 
      \frac{1}{R} D_{l, m}^{\left(l_1, m_1\right), \left(l_2, m_2\right)} 
      \dot a_{l_1 , m_1} \dot a_{l_2 , m_2} \\
      & + \frac{\dot R}{R^2} Z_{l, m}^{\left(l_1, m_1\right), \left(l_2, m_2\right)} 
      \dot a_{l_1 , m_1} a_{l_2 , m_2}
      + \biggl[\frac{\ddot R}{R^2} K_{l, m}^{\left(l_1, m_1\right), \left(l_2, m_2\right)} \\
      & + \frac{\alpha}{R^4} C_{l, m}^{\left(l_1, m_1\right), \left(l_2, m_2\right)} +
      \frac{\dot{R}^2}{R^3} X_{l, m}^{\left(l_1, m_1\right), \left(l_2, m_2\right)}
      \biggr] a_{l_1 , m_1} a_{l_2 , m_2}
      ,
  \end{aligned}
\end{gather}

\begin{gather}
  \label{eq: radius eom}
\begin{aligned}
  R\ddot{R} = 
  \Biggl\{ & -\frac{3}{2}\dot{R}^{2}-\Xi-\frac{2\alpha}{R}
  \frac{1}{4\pi}\sum_{lm}
  \biggl[\frac{2l+1}{2\left(l+1\right)}|\dot{a_{lm}}|^{2} \\
  & + \frac{l - 1}{l + 1}\frac{\dot{R}}{R} \dot{a_{lm}}a_{lm}^{*}
  +\left(
      \frac{l - 1}{l + 1} \frac{\dot{R} ^2}{R^2}
      +\left(3l - l^3\right)\frac{\alpha}{R^{3}}
  \right)|a_{lm}|^{2}\biggr] \Biggr\} \\
  & / \left\{1-\frac{1}{4\pi}\frac{1}{R^{2}}
  \sum_{lm} l |a_{lm}|^{2} \right\}
  .
\end{aligned}
\end{gather}

\begin{figure*}[!t]
  \begin{minipage}{0.45\linewidth}
  \begin{subfigure}{\textwidth}
    \includegraphics[width=\linewidth]{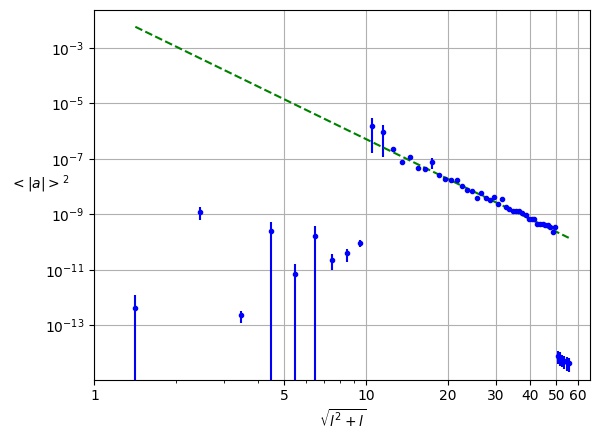}
    \caption{} 
    \label{fig:spectrum_avg}
  \end{subfigure}
  \end{minipage}
  \begin{minipage}{0.45\linewidth}
  \begin{subfigure}{\textwidth}
    \includegraphics[width=\linewidth]{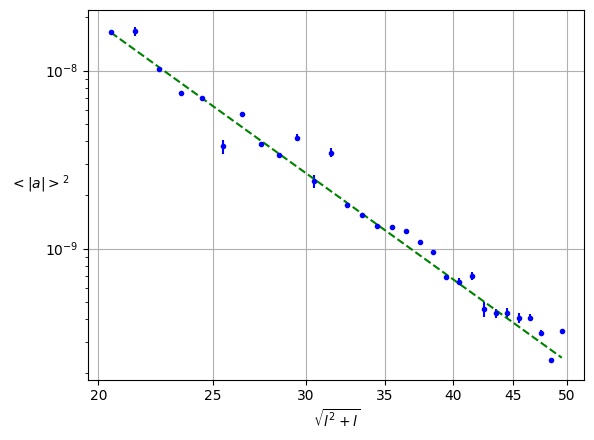}
    \caption{} 
    \label{fig:spectrum_avg_narrow}
  \end{subfigure}
  \end{minipage}

\caption{
An example of the mean squared elevation. 
The entire domain appears in (a).
In (b), only the interval predicted inertial, between pumping and the onset of high mode dissipation.
Dashed green lines denote fits and blue dots simulation data. 
Data is fitted in the range $ l \in \left[20, 35\right]$.
This simulation was run with $\beta = 2.1$, $\epsilon = 0.1$ and $l_{max} = 55$.
Other parameters as previously defined.} 
\label{fig:spectrum}
\end{figure*}

Einstein's summation notation is applied in \eqref{eq: perturbation eom}.
$A_l$ corresponds to $-\omega_{l} ^2 = - W_l$ defined in the introduction.
It is a private case of the result by Plesset
\cite{plesset_stability_1954} for empty bubbles
and defined in \eqref{eq: A definition}. 

Interaction tensors appear in the appendix \eqref{eq: interaction tensors} 
and are proportional to the Gaunt coefficients with some rational, unitless factor of $l_1 ,l_2$ and $l$,
as a result of curved, spherical geometry.
In wave turbulence theory, interaction is proportional to wave overlap.
The Hamiltonian is a scalar, integral quantity.
Since eigenmodes of the linear Hamiltonian accompany amplitude,
the Hamiltonian term $\mathcal{H}_j$ describing $j$-wave interaction
contains terms proportional to $j$-wave overlap.
In flat geometry, harmonics diagonalize the linear Hamiltonian.
Their overlap is simply
$\delta\left(\sum \vec k_{in} - \sum \vec k_{out}\right)$.
In our case, eigenmodes are the spherical harmonics.
The $3$-wave overlap is therefore proportional to the Gaunt coefficients.
These also appear as interaction coefficients in \cite{zhang_weakly_2018},
though in a lower dimension.

The series appearing in the equation for $\ddot R$ 
do not necessarily converge.
Suggesting a power-law $a_l \sim l^{-\beta}$,
we must require that $\beta > 2$, 
as shown in \ref{subsection: initial conditions limit}.
For amplitudes to display non-power-law behavior,
one must define another length scale between pumping and dissipation scales.
Therefore, we disregard it as an option for initial conditions.
Note that the theoretical spectrum satisfies this demand.

The equations are numerically solved as described in appendix \ref{appendix: schemes}.
Our code is publically available\footnote{
  Git repository:
  \href{https://gitlab.com/pelegemanuel/empty_cavity}{empty cavity}.
}.

We add artificial pumping and dissipation terms following previous works, primarily \cite{pushkarev_turbulence_1996}.
Exact forms and parameters appear in appendix
\ref{section: artificial}.
One should note that dissipation is required 
in both long and short wavelengths. 
The former for spectra emergence, the latter for bubble stability,
since low $l$'s might become relatively large.
While this makes a further requirement on artificial terms, 
it is important to note that we don't add next order terms, 
required in \cite{pushkarev_turbulence_1996} for stability.

We run all simulations with 
$R_0 = 1$ and $\alpha = 1$, setting a time scale
$\sqrt{R_0 ^3 / \alpha}$.
In addition, 
we set $P_\infty = 0$, $V_0 = \frac{4\pi}{3} R_0 ^3$,
and $P_0$ such that the initial unperturbed bubble is at equilibrium.
Other parameters appear in the appendix.
The parameters $\beta$, 
maximal angular momentum number $l_{max}$, 
and $\epsilon$, a linear scale of the initial perturbations,
differ among simulations.

Interaction tensor multiplication is by far
the most demanding task numerically,
even when parallelization, sparse matrix algorithms,
and designated libraries are applied.
We therefore approximate nonlinear terms, 
as detailed in appendix \ref{subsection: interaction approximation schemes}.
First, by extrapolation.
Second, by ignoring sub-leading terms, 
which, although less relevant for dynamics,
are just as demanding computationally.

\begin{figure}[!t]
  \center
  \captionsetup{width=\linewidth}
  \includegraphics[width=\linewidth]{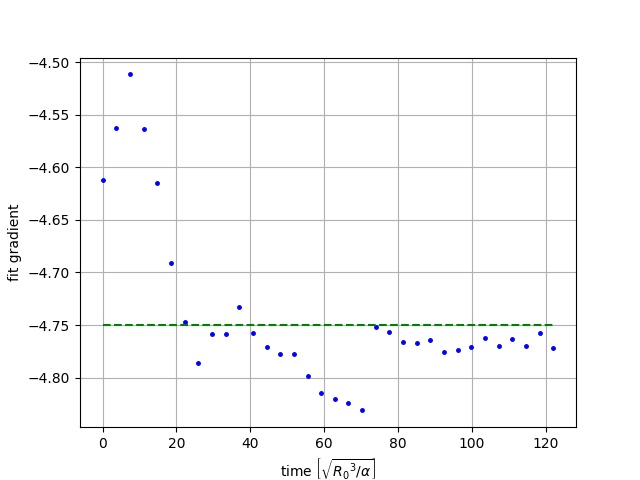}
  \caption{
  Fitted power law as a function of time.
  Simulation parameters were selected as in figure \ref{fig:spectrum}.
  } 
  \label{fig:spectrum_fit}
\end{figure}

\begin{figure}[!b]
  \includegraphics[width=\linewidth]{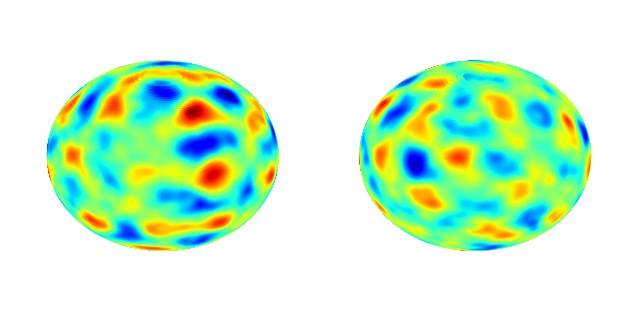}
  \caption{
  An example state of the bubble, projected on a globe
  (from both sides).
  Simulation parameters were selected as in figure \ref{fig:spectrum}.} \label{fig:elevation}
\end{figure}

Figure \ref{fig:spectrum} shows an example
average elevation spectrum the system arrives at
for when the fitted power-law is stable.
Height maps for bubbles of the same simulation appear in
figure \ref{fig:elevation}.

By looking at \ref{fig:spectrum_avg}, one might suspect that
the inertial interval is $\simeq [20,50]$. 
However, a closer numerical inspection suggests the range $[20,35]$.
Pumping and high mode dissipation set the bounds for the inertial interval.
Dissipation starts sharply at $l = 50$, and pumping is centered around $l = 15$,
with short decay length, bounding the inertial interval in $[20,50]$.
However, there is no guarantee that interactions for all modes in this range are simulated effectively since $l_{disp}$ limits possible interactions.
The decay of modes with $l$ promises that for the dynamics of a mode with $l \ll l_{disp}$,
modes above $l_{disp}$ are negligible.
Nevertheless, approaching $l_{disp}$, 
we cannot disregard the missing contributions in such a manner.
As a result, the inertial interval is cut short before $l_{disp}$.
We are computationally limited to $l_{max} \le 70$, 
and choose $l_{disp} = 50$, leaving a rather narrow dissipation range.
Still, this doesn't allow us to simulate interactions effectively in a wide band. 
We numerically find that in the range $[20, 35]$ 
the spectrum changes rather quickly. 
When fitting a wider interval, a change in spectrum
is slower, if evident at all.
Figure \ref{fig:spectrum_fit}
presents the fitted power-law at different times.
It quickly converges and stabilizes in the vicinity of the theoretical $-4.75$
\cite{zakharov_weak_1971,zakharov_kolmogorov_2012}.

In a low gravity experiment conducted on a similar system,
elevation amplitudes presented hexagonal patterns
when forcing was periodic \cite{falcon_capillary_2009}. 
We have been able to trace hints of such patterns in our results.
Figure \ref{fig:elevation} shows a simulated bubble surface
for example. 
Looking at the left globe,
two overlapping hexagonal patterns seem to cover 
most of the visible side of the globe.
The obtained pattern wavelength 
is much larger than its experimental value. 
In \cite{falcon_capillary_2009}, 
an inertial interval
of two decades was measured. 
Our simulation is computationally limited,
and figure \ref{fig:elevation}
was extracted from a simulation with an inertial interval of less
than $0.4$ decades, at lower wavenumbers,
explaining the appearance of rougher patterns.
This observation of hexagonal pattern formation
is not conclusive and remains to be further studied.

%% file: paperDiscussion.tex
\section{Discussion}
The main contribution of this work
is the establishment of a connection
between two branches in the field of hydrodynamic instabilities,
wave turbulence theory and bubble dynamics.
The field of hydrodynamic instabilities is vast,
studied analytically, experimentally, and predominantly numerically.
Since simulations are limited, researchers often turn to simplified
systems, focusing on certain aspects of systems at the expense of others.
For example, nonlinear Rayleigh-Tayor instabilities
are often simulated in two dimensions
and under limitations on wavelengths
\cite{wang_weakly_2015, zhang_weakly_2018, zhao_two-dimensional_2019}.

The wave turbulence framework contains extensive tools and predictions.
By connecting it and bubble perturbations, 
we hope to apply these methods to the study of hydrodynamic instabilities.
With this purpose in mind,
we move forward to discuss the limitations of this study.
Overcoming each of them constitutes a direction
for future work, 
possibly extending the validity of our results
to either 
gravity waves, unstable bubbles, or other geometries. 

First is the simulation of acceleration-controlled turbulence.
Such simulation is relevant for studies of cavitation and sonoluminescence, in which bubble surfaces are subjected to immense accelerations.
Weak turbulence theory predicts that gravity wave turbulence is $4$-wave dominant,
requiring the simulation of next-order terms,
thus posing a computational challenge.
The costly calculations of $3$-wave interactions
constrained our simulation, taking almost all running time and memory.
Inclusion of $4$-wave interaction is supposed to
either strongly limit $l_{max}$
or raise running time by several orders of magnitude.

Second is the study of rapid phenomena,
of which cavitation and sonoluminescence are also exemplary.
Many predictions of wave turbulence theory,
including the Kolmogorov-Zakharov spectrum corroborated here,
are derived assuming
turbulence is fully developed and stationary.
Rapid phenomena challenge this assumption,
since turbulence does not necessarily have the time
to develop and stabilize, compromising the validity of our theory.

Lastly,
the effects of large wavelength perturbations
and curvature on turbulent phenomena
are only partially addressed in this work.
Neither our analytical nor our numerical
results hold for them.
The analytical,
since the analogy presented in the linear regime only
applies for small wavelengths. 
The numerical, since dissipation suppresses large wavelengths in our simulation. 
While large wavelength contributions
are subject to countless numerical studies, 
possible small-large wavelength interactions
remain to be investigated.
In addition, 
a more general, perhaps universal, approach to
the study of curvature effects on turbulence is currently absent.  

To conclude,
this study links the fields of wave turbulence and bubble dynamics. 
We were able to simulate capillary wave turbulence
on the surface of a stable bubble.
Our results are in agreement with
the Kolmogorov Zakharov spectrum predicted by weak turbulence theory 
and verified in low gravity experiments.
We hope that the extensive tools developed for wave turbulence
will be used for the study of bubble dynamics.

%% file: paperAppendixGaunt.tex
\section{Spherical harmonics and Gaunt coefficients}
The nonlinear wave framework refers to the eigenfunctions of the
linear Hamiltonian as waves. Usually, these are the regular harmonics.
However, working in the curved, spherical system, 
our waves are the spherical harmonics, 
as found by Plesset \cite{plesset_stability_1954}.
We use the normalization

\begin{equation}
    Y_{l,m} \left(\Omega\right) = \left( - 1\right)^{ - m}
    \sqrt{\frac{2l + 1}{4\pi} \frac{\left(l - m\right)!}{\left(l + m\right)!}}
    P_{l,m} \left(\cos \theta \right)
    e^{i m \phi}
    .
\end{equation}

Here $P_{lm}$ are the associated Legendre polynomials. 
Spherical harmonics are defined for every 
$l \in \mathbb{N}$ and integer $m$ such that $-l \le m \le l$.
They form an orthonormal set, namely 
$\int d\Omega Y_{l,m} Y_{l',m'}^* = \delta_{l, l'} \delta_{m, m'}$ 
with $\delta$ the Kronecker delta.
Non linear perturbation theory requires that
we find $\mathcal{H}_j$ for $j > 2$. As each
perturbation amplitude is accompanied by a spherical harmonic,
terms in $\mathcal{H}_j$ contain the integral of $j$
spherical harmonics overlap. In leading nonlinear order, $j = 3$,
and such overlap is the definition of the Gaunt coefficients,

\begin{equation}
    G^{m, m_1 , m_2}_{l, l_1 , l_2} = \int d\Omega Y_{l,m} 
    \left(\Omega\right) Y_{l_1 ,m_ 1} \left(\Omega\right) Y_{l_2 ,m_2} \left(\Omega\right)
    .
\end{equation}

These coefficients are tightly related to the Wigner $3j$ symbols and the Clebsch-Gordon coefficients.
For some intuition, consider the quantum analogy discussed earlier,
in which $\vec k$ conservation corresponds to momentum conservation.
In spherical geometry, the relevant conservation law is that of angular momentum.
$2 \to 1$ interaction would thus require that we move to the total
angular momentum frame for the two incoming waves to find that of the outgoing one,
explaining the appearance of Clebsch-Gordon-like coefficients.
It is important to note that this analogy is not an explanation, 
but rather an outcome of a similar mathematical representation. 

Besides the Gaunt coefficient, we may encounter several more
integrals containing spatial contravariant derivatives of spherical harmonics.
Denote by $\vec \nabla _{\Omega}$ the angular part of the gradient in spherical coordinates.
Recalling that $\nabla^2 _{\Omega} Y_{lm} = -l \left(l+1\right) Y_{lm}$,
the only other, non-trivial integral we encounter in our derivation is
$\int d\Omega Y \vec \nabla Y \cdot \vec \nabla Y$.
This expression is, in fact, also proportional to a Gaunt coefficient. 
Integrating by parts, we define the following $\epsilon_{l,l_1,l_2}$ via

\begin{gather}
\begin{aligned}
    \int d\Omega Y_{l,m} \left(\Omega\right)
    \vec \nabla  Y_{l_1 ,m_1} \left(\Omega\right) 
    \cdot
    \vec \nabla  Y_{l_2 ,m_2} \left(\Omega\right) 
    & = \epsilon_{l,l_{1},l_{2}} G^{m, m_1 , m_2}_{l, l_1 , l_2} \\
    & = \frac{1}{2}\left[l_{1}(l_{1}+1)+l_{2}(l_{2}+1)-l(l+1)\right] G^{m, m_1 , m_2}_{l, l_1 , l_2}
    . 
\end{aligned}
\end{gather}

We conclude that all interaction coefficients are proportional to Gaunt coefficients.
Note that tensors comprised of such coefficients are sparse, 
as Gaunt coefficients are non-zero only when the following conditions hold. 

\begin{itemize}
    \label{gauntConditions}
    \item $m + m_1 + m_2 = 0$
    \item $l + l_1 + l_2 = 0 \mod 2$
    \item $ |l_1 - l_2| \le l \le l_1 + l_2$
\end{itemize}

The first two conditions arise from parity. The third is the triangle inequality. 

%% file: paperAppendixGeometry.tex
\section{Geometry of semispheres}
\label{section: semisphere geometry}

\subsection{Surface and normal calculations}

Let $\psi\left(\Omega\right)$ be a local radius scalar field.
We define a semisphere by the following set of points,
represented here in the standard basis of $\mathbb{R}^3$,

\begin{gather}
    \vec p = 
    \begin{pmatrix}
        \psi\left(\theta, \phi\right) \sin \theta \cos \phi \\
        \psi\left(\theta, \phi\right) \sin \theta \sin \phi \\
        \psi\left(\theta, \phi\right) \cos \theta
    \end{pmatrix}
    .
\end{gather}

Using this exterior geometry representation,
we derive vector surface components, as well as the metric,

\begin{gather}
    g_{ij} = 
    \begin{pmatrix}
        \psi^2 + \left(\partial_{\theta} \psi \right)^2 &
        \partial_{\theta} \psi \partial_{\phi} \psi \\
        \partial_{\theta} \psi \partial_{\phi} \psi &
        \psi^2 \sin^2 \theta + \left(\partial_{\phi} \psi \right)^2
    \end{pmatrix}
    , 
\end{gather}

\begin{gather}
    \label{eq:normal vec}
    \vec {dS} = \psi^2 \sin\theta \left(\hat r
    - \frac{1}{\psi} \vec \nabla_{\Omega} \psi \right)
     = \psi^2 \sin\theta \sqrt{1 +\left(\frac{\vec \nabla_{\Omega}\psi}{\psi}\right)^2}
     \hat n
     .
\end{gather}

\subsection{Principle radii}
The principle radii are computed following
\cite[section 275]{horace_lamb_hydrodynamics_1916}. 
Suppose our surface is defined by some scalar field
$F\left(\vec r\right)$, such that $F > 0$ outside,
$F = 0$ on the surface, and $F < 0$ inside,
then

\begin{gather}
    \frac{1}{R_1} + \frac{1}{R_2} = \vec \nabla \cdot 
    \left(\frac{1}{|\vec \nabla F|} \vec \nabla F \right)
    .
\end{gather}

Note that for semispheres
$F\left(\vec r\right) = r - \psi\left(\theta, \phi\right)$. 
It is useful to note that $\partial_{i} \psi$
is spanned by $\hat \theta$ and $\hat \phi$ and thus $x^i \partial_{i} \psi = 0$.
We get

\begin{gather}
    \frac{1}{R_{1}}+\frac{1}{R_{2}}
    =\frac{1}{\sqrt{1 + \partial_{k} \psi \partial^{k} \psi}}
    \left[\nabla^{2}F-\frac{1}{2}
    \frac{\partial^{j}
    \left(1 + \partial_{i} \psi \partial^{i} \psi\right)}
    {1 + \partial_{l} \psi \partial^{l} \psi}
    \left(x_j-\partial_{j} \psi\right)\right]
    .
\end{gather}

Noting that

\begin{gather}
    x^j \partial_j \partial_{i} \psi = \partial_j 
    \left(x^j \partial_{i} \psi\right) - \partial_{i} \psi = - \partial_{i} \psi
\end{gather}

and

\begin{gather}
    \nabla^{2}F = \frac{2}{r}-\frac{1}{r^{2}}\nabla_{\Omega}^{2}\psi
    ,
\end{gather}

we finally arrive at

\begin{gather}
\begin{aligned}
    \frac{1}{R_{1}}+\frac{1}{R_{2}} & = 
    \frac{1}{\left(1+\vec{\nabla}\psi\cdot\vec{\nabla}\psi\right)^{3/2}} \\
    & \times
    \left[\left(1+\vec{\nabla}\psi\cdot\vec{\nabla}\psi\right)
    \left(\frac{2}{\psi}-\nabla^{2}\psi\right)
    +\left(\frac{1}{\psi}\vec{\nabla}\psi\cdot\vec{\nabla}\psi
    +\frac{1}{2}\vec{\nabla}\left(\vec{\nabla}\psi\cdot\vec{\nabla}\psi\right)
    \cdot\vec{\nabla}\psi\right)\right]
    .
\end{aligned}
\end{gather}

%% file: paperAppendixDerivation.tex
\section{Derivation}

We now move forward to derive our equations of motion
by plugging previous results and series expansions in Bernoulli's equation.
We start by expanding geometrical quantities.
Then, we use the volumetric flux equation to connect the 
perturbation amplitudes and velocities to the fluid velocity coefficients $b_{lm}$.
Having done that, 
we continue to get our second-order equations of motion by plugging all terms
in Bernoulli's condition.

It is important to note that perturbation amplitudes may be complex,
as spherical harmonics are complex. However, the local radius must always be real.
Requiring that $\psi^* = \psi$, and recalling that spherical harmonics form
a linearly independent set and that $Y_{l,m}^* = \left(-1\right)^m Y_{l,-m}$, we
find

\begin{equation}
    \label{eq: realness condition}
    a_{l, - m} = \left(-1\right)^m a_{l,m}^*
    .
\end{equation}

This equation should be respected by both our initial conditions and the
derived equations of motion.

\subsection{Geometry}
The expression for the harmonic mean of the principal radii 
is greatly simplified in second-order.
The last term is of leading third-order, 
and can be ignored. We get

\begin{gather}
    \label{eq:principle radii}
\begin{aligned}
    \frac{1}{R_{1}}+\frac{1}{R_{2}}
    & = \frac{2}{\psi}-\frac{1}{\psi^{2}}\nabla_{\Omega}^{2}\psi \\
    & = \frac{2}{R}
    +\frac{1}{R^2}\sum_{l,m}\left(l^{2}+l-2\right)a_{lm}Y_{lm}
    -\frac{1}{R^3}\sum_{l,m,l',m'}2\left(l^{2}+l-1\right)a_{lm}a_{l'm'}Y_{lm}Y_{l'm'}
    .
\end{aligned}
\end{gather}

In first-order we recreate the result by Plesset \cite{plesset_stability_1954}.
We calculate the volume using the realness condition \eqref{eq: realness condition},
and find

\begin{equation}
    V = \int d\Omega \int_0^{\psi} r^2 dr 
    = \frac{1}{3} \int d\Omega \psi^3
    = \frac{4\pi}{3} R^3 + R \sum_{l,m} |a_{l, m}|^2
    + O\left(a^3\right)
    .
\end{equation}

\subsection{The fluid velocity coefficients}
Plugging the surface vector we've derived \eqref{eq:normal vec}
into the volumetric flux equation, we get

\begin{equation}
    v_{r}(\psi)-\dot{\psi}
    -\frac{1}{\psi} \vec{\nabla}_{\Omega}\psi \cdot \vec{v_{\Omega}} = 0
\end{equation}

where $v_r$ and $\vec{v_{\Omega}}$ the radial and angular parts of the fluid velocity, respectively.
Note that both $\vec{v_{\Omega}}$ and $\vec{\nabla}_{\Omega}\psi$ are of leading
first order in the perturbation. 
As a result, in first-order, the last term disappears and
we find $v_r = \dot \psi$ as suggested by Plesset, hence

\begin{equation}
    b_{lm}^{(1)}=\dot{a_{lm}}+2\frac{\dot{R}}{R}a_{lm}
    .
\end{equation}

We use upper indices in brackets to denote the order of the term in perturbation amplitudes and velocities.
Up to second order, $b_{lm} \sim b_{lm}^{(1)} + b_{lm}^{(2)}$. 
Plugging this into our flux equation,
keeping only terms up to second order, one finds

\begin{equation}
    b_{lm}^{(2)} =
    (-1)^{m}\sum G_{l,l_{1},l_{2}}^{-m,m_{1},m_{2}}
    \left(g_{l,l_{1},l_{2}}
    \frac{1}{R}b_{l_{1}m_{1}}^{(1)}a_{l_{2}m_{2}}
    -3\frac{\dot{R}}{R^{2}}a_{l_{1}m_{1}}a_{l_{2}m_{2}}\right) 
\end{equation}

where

\begin{equation}
    g_{l,l_{1},l_{2}} \equiv l_{1} + 2
    -\frac{\epsilon_{l,l_{1},l_{2}}}{l_{1} + 1}
    .
\end{equation}

Higher-order expressions for $b$ may be obtained by repeating the above process.

\subsection{First order equations of motion}

Up to second-order, our equation of motion takes the form

\begin{gather}
    \Xi + \frac{\partial \phi}{\partial t} - \frac{1}{2} v^2 
    + \alpha \left(\frac{2}{\psi}-\frac{1}{\psi^{2}}\nabla_{\Omega}^{2}\psi\right) = 0
    .
\end{gather}

In first-order, the velocity of the fluid is radial. Hence,

\begin{gather}
\begin{aligned}
    & \left(\frac{\partial_{t}\left(\dot{R}R^{2}\right)}{R}
    \left(1-\frac{1}{R}\sum_{lm}a_{lm}Y_{lm}\right)
    +\sum_{lm}\left(\frac{l+2}{l+1}\dot{R}b_{lm}
    -\frac{1}{l+1}R\dot{b_{lm}}\right)Y_{lm}\right) \\
    & -\frac{1}{2}\left(\dot{R}^{2}
    -4\dot{R}^{2}\frac{1}{R}\sum_{lm}a_{lm}Y_{lm}+2\dot{R}\sum_{lm}b_{lm}Y_{lm}\right)
    +\Xi + \alpha\left(\frac{2}{R}+\frac{1}{R}\sum\left(l^{2}+l-2\right)a_{lm}Y_{lm}\right)
    = 0
    .
\end{aligned}
\end{gather}

Simple integration yields the Rayleigh-Plesset equation

\begin{equation}
    R\ddot{R}+\frac{3}{2}\dot{R}^{2}+\Xi + \frac{2\alpha}{R} = 0
    .
\end{equation}

If we first multiply by some $Y_{lm}^*$, we find the perturbation equation

\begin{equation}
    \frac{1}{l + 1}R\ddot{a}_{lm}
    +\frac{3}{l + 1}\dot{R}\dot{a}_{lm}
    -\left[\frac{l - 1}{l + 1}\ddot{R}
    -\frac{\alpha}{R^{2}}\left(l+2\right)\left(l-1\right)\right]a_{lm}= 0
    .
\end{equation}

Denote by $q_l \equiv \left(l+2\right)\left(l+1\right)\left(l-1\right)$.
We find, after multiplying by $\left(l + 1\right)$,

\begin{equation}
    \label{eq: first order equation}
    \ddot{a}_{lm}+3\frac{\dot{R}}{R}\dot{a}_{lm}-A_{l}a_{lm}= 0
    ,
\end{equation}

where

\begin{equation}
    \label{eq: A definition}
    A_{l} \equiv \left(l-1\right)\frac{\ddot{R}}{R}
    -\frac{\alpha}{R^{3}}q_l
    .
\end{equation}

It is important to note that 
we recreate the result by Plesset\cite{plesset_stability_1954}
for empty cavities.
This is also in agreement with the spectrum of capillary-gravity waves.
Note that Fourier modes 
$e^{-i\vec k \cdot \vec r}$ are eigenmodes of 
$\nabla^2$ with eigenvalues $-k^2$.
$Y_{lm}$ are eigenmodes of 
$\nabla_{\Omega} ^2$ with eigenvalues $-l\left(l+1\right)$.
For $l \gg 1$ we can thus suggest $k \sim l / R$. Now

\begin{equation}
    \label{eq: plesset zakharov correspondence}
    A_l \sim - \left(\sqrt{\left( - \ddot R \right) k + \alpha k^3}\right)^2 
    = -\omega^2 \left(k ; - \ddot R, \alpha \right)
    .
\end{equation}

Where $\omega \left(k ; g, \alpha \right)$ is the deep water wave angular frequency for wavelength $k$,
gravity constant $g$ and capillarity constant $\alpha$. 
For small wavelengths ($l \gg 1$), $\omega \gg \frac{\dot{R}}{R}$,
we can ignore the second term and see that small waves follow deep water wave theory.
The second term adds friction/pumping, 
depending on the direction in which the mean radius moves.
Either way, for $l \gg 1$, the change in wave amplitude it generates happens at a much longer time scale.

\subsection{Second order equations of motion}

We should add to our equations the second-order component,
given by the following formula

\begin{equation}
    s.o.c
    =\dot{\phi}^{(2)}-\frac{1}{2}(v^{2})^{(2)}
    +\alpha\left(\frac{1}{R_{1}}+\frac{1}{R_{2}}\right)^{(2)}
    .
\end{equation}

Having already calculated the last term, we turn to the first two,
and find

\begin{gather}
\begin{aligned}
    \dot{\phi}^{(2)} |_{r=\psi}
    & = \left[R\ddot{R}+2\dot{R}^{2}\right]
    \left(\sum_{lm}\frac{1}{R}a_{lm}Y_{lm}\right)^{2}
     + \sum_{lm} Y_{lm}\left(\frac{1}{l+1}\left(\dot{b_{lm}^{(2)}}R+(l+2)b_{lm}^{(2)}\dot{R}\right)\right) \\
    & -\frac{1}{R}\sum_{lm} Y_{lm}Y_{l'm'}\left(\frac{1}{R}a_{l'm'}\left(\dot{b_{lm}^{(1)}}R+(l+2)b_{lm}^{(1)}\dot{R}\right)\right)
    ,
\end{aligned}
\end{gather}

\begin{gather}
\begin{aligned}
    \left(v^{2}\right)^{(2)} |_{r=\psi}
    & = 2v_{r}^{(0)}v_{r}^{(2)}+v_{r}^{(1)}v_{r}^{(1)}
    + v_{\Omega}^{(1)}v_{\Omega}^{(1)}|_{r=\psi} \\ 
    & = 2\dot{R}\sum_{lm}b_{lm}^{(2)}Y_{lm} \\
    & +\sum_{lml'm'}Y_{lm}Y_{l'm'}\left(\dot{a_{lm}}\dot{a_{l'm'}}
    -2(2l+1)\frac{\dot{R}^{2}}{R^{2}}a_{lm}a_{l'm'}
    -2(l+2)\frac{\dot{R}}{R}\dot{a}_{lm}a_{l'm'}\right) \\
    & +\left(\sum_{lm}\frac{1}{l+1}b_{lm}\vec{\nabla}Y_{lm}\right)^{2}
    .
\end{aligned}
\end{gather}

Combining these, we get

\begin{gather}
\begin{aligned}
    s.o.c & =\sum_{lm} Y_{lm}\left(\frac{1}{l+1}
    \left(\dot{b_{lm}^{(2)}}R+b_{lm}^{(2)}\dot{R}\right)\right)\\
    & -\sum_{lml'm'}Y_{lm}Y_{l'm'}
    \left(
        \frac{1}{2}\dot{a_{lm}}\dot{a_{l'm'}}
        -\frac{\dot{R}}{R}\dot{a_{lm}}a_{l'm'}
    +\left[
        l\frac{\ddot{R}}{R}
        - \frac{\dot{R}^{2}}{R^{2}}
        + \left(3l - l^3\right) \frac{\alpha}{R^3}
    \right]
    a_{lm}a_{l'm'}\right)\\
    & -\frac{1}{2}\sum_{lml'm'}\frac{1}{(l+1)(l'+1)}b_{lm}b_{l'm'}\left(\vec{\nabla}Y_{lm}\cdot\vec{\nabla}Y_{l'm'}\right)
    .
\end{aligned}
\end{gather}

Here we've plugged in the first order acceleration in 
$\dot b^{(1)}$ terms. We can now
derive the equations governing radius motion, 
together with second-order corrections,

\begin{gather}
\begin{aligned}
    4\pi & \left(R\ddot{R}+\frac{3}{2}\dot{R}^{2}
    +\Xi+\frac{2\alpha}{R}\right) \\
    & =\sum_{lm} \left(-1\right)^{m}
    \left(\frac{2l+1}{2\left(l+1\right)}\dot{a_{lm}}\dot{a_{l-m}}
    +\frac{l-1}{l+1}\frac{\dot{R}}{R}\dot{a_{lm}}a_{l-m}
    +\left[
        l\frac{\ddot{R}}{R}
        +\frac{l - 1}{l + 1}\frac{\dot{R}^{2}}{R^{2}}
        + \left(3l - l^3\right) \frac{\alpha}{R^3}
    \right]a_{lm}a_{l-m}\right)
    .
\end{aligned}
\end{gather}

Note that $\ddot R$ appears on both sides. 
Moving all of its occurrences to the left,
together with the realness condition
\eqref{eq: realness condition} 
finish the derivation
of the radial equation of motion

\begin{gather}
\begin{aligned}
    R\ddot{R} & = 
    \Biggl\{ -\frac{3}{2}\dot{R}^{2}-\Xi-\frac{2\alpha}{R} \\
    & + \frac{1}{4\pi}\sum_{lm}
    \biggl[\frac{2l+1}{2\left(l+1\right)}|\dot{a_{lm}}|^{2}
    +\frac{l - 1}{l + 1}\frac{\dot{R}}{R} \dot{a_{lm}}a_{lm}^{*}
    +\left(
        \frac{l - 1}{l + 1} \frac{\dot{R} ^2}{R^2}
        +\left(3l - l^3\right)\frac{\alpha}{R^{3}}
    \right)|a_{lm}|^{2}\biggr] \Biggr\} \\
    & / \left\{1-\frac{1}{4\pi}\frac{1}{R^{2}}
    \sum_{lm} l |a_{lm}|^{2} \right\}
    .
\end{aligned}
\end{gather}

As for the perturbation second-order equations, 
we integrate after multiplying by some $Y_{lm} ^*$ to find
the following equation. We multiply by $l+1$,
to match the only operation we did 
in the derivation of the first-order equation
\eqref{eq: first order equation}.

\begin{gather}
\begin{aligned}
    \ddot{a}_{lm} &+3\frac{\dot{R}}{R}\dot{a}_{lm}-A_{l}a_{lm} =
    -\frac{1}{R}\left(\dot{b_{lm}^{(2)}}R+b_{lm}^{(2)}\dot{R}\right)
     +(-1)^{m}\frac{l + 1}{R}
    \sum_{l_{1}m_{1}l_{2}m_{2}}G_{l,l_{1},l_{2}}^{-m,m_{1},m_{2}}
    \Biggl[
    \frac{1}{2}\dot{a_{l_{1}m_{1}}}\dot{a_{l_{2}m_{2}}} \\
    & - \frac{\dot{R}}{R}\dot{a}_{l_{1}m_{1}}a_{l_{2}m_{2}}
    +\left(
        l_1 \frac{\ddot{R}}{R}
        - \frac{1}{2}\frac{\dot{R}^{2}}{R^{2}}
        + \left(3 l_1 - {l_1}^3\right) \frac{\alpha}{R^3}
    \right)
    a_{l_{1}m_{1}}a_{l_{2}m_{2}} 
    \Biggr]\\
    & +(-1)^{m}\frac{l + 1}{R}\sum_{l_{1}m_{1}l_{2}m_{2}}
    \frac{\epsilon_{l,l_{1},l_{2}}}{2(l_{1}+1)(l_{2}+1)}G_{l,l_{1},l_{2}}^{-m,m_{1},m_{2}}
    \left(\dot{a}_{l_{1}m_{1}}\dot{a}_{l_{2}m_{2}}
    +4\frac{\dot{R}}{R}\dot{a}_{l_{1}m_{1}}a_{l_{2}m_{2}}
    +4\frac{\dot{R}^{2}}{R^{2}}a_{l_{1}m_{1}}a_{l_{2}m_{2}}\right)
    .
\end{aligned}
\end{gather}

Note that

\begin{gather}
\begin{aligned}
    \partial_{t}(Rb_{lm}^{(2)}) = (-1)^{m}\sum G_{l,l_{1},l_{2}}^{-m,m_{1},m_{2}}
    \Biggl\{ &\left[g_{l,l_{1},l_{2}}A_{l_{1}}
    +\left(2g_{l,l_{1},l_{2}}-3\right)\left(\frac{\ddot{R}}{R}
    -\frac{\dot{R}^{2}}{R^{2}}\right)\right]a_{l_{1}m_{1}}a_{l_{2}m_{2}}\\
    +&\left(-g_{l,l_{1},l_{2}}+2g_{l,l_{2},l_{1}}-6\right)
    \frac{\dot{R}}{R}\dot{a}_{l_{1}m_{1}}a_{l_{2}m_{2}}
    +g_{l,l_{1},l_{2}}\dot{a_{l_{1}m_{1}}}\dot{a_{l_{2}m_{2}}}
    \Biggr\}
    .
\end{aligned}
\end{gather}

All that is left now is to sum together terms with the different coefficients. We get

\begin{gather}
    \begin{aligned}
        \ddot a_{l,m} + 3 \frac{\dot R}{R} \dot a_{l,m} - A_{l} a_{l,m} & = 
        \left[\frac{\ddot R}{R^2} K_{l, m}^{\left(l_1, m_1\right), \left(l_2, m_2\right)} +
        \frac{\alpha}{R^4} C_{l, m}^{\left(l_1, m_1\right), \left(l_2, m_2\right)} +
        \frac{\dot{R}^2}{R^3} X_{l, m}^{\left(l_1, m_1\right), \left(l_2, m_2\right)}
        \right] a_{l_1 , m_1} a_{l_2 , m_2} \\
        & + \frac{\dot R}{R^2} Z_{l, m}^{\left(l_1, m_1\right), \left(l_2, m_2\right)} 
        \dot a_{l_1 , m_1} a_{l_2 , m_2}
        + \frac{1}{R} D_{l, m}^{\left(l_1, m_1\right), \left(l_2, m_2\right)} 
        \dot a_{l_1 , m_1} \dot a_{l_2 , m_2}
        ,
    \end{aligned}
\end{gather}

where

\begin{gather}
    \label{eq: interaction tensors}
    \begin{aligned}
K_{l,l_{1},l_{2}}^{m,m_{1},m_{2}} & =
(-1)^{m}G_{l,l_{1},l_{2}}^{-m,m_{1},m_{2}}
\biggl[
    l_1 \left(l + 1\right) 
    - \left(l_1 - 3\right)g_{l,l_1,l_2} 
    + 3
\biggr]
\\ C_{l,l_{1},l_{2}}^{m,m_{1},m_{2}} & =
(-1)^{m}G_{l,l_{1},l_{2}}^{-m,m_{1},m_{2}}
\biggl[
    \left(l + 1\right) \left(3 l_1 - l_1^3\right) 
    + g_{l, l_1, l_2} q_{l_1}
\biggr]
\\ X_{l,l_{1},l_{2}}^{m,m_{1},m_{2}} & = 
(-1)^{m}G_{l,l_{1},l_{2}}^{-m,m_{1},m_{2}}
\left[ 
    - l - 4 
    + 2\left(l + 1\right)
    \frac{\epsilon_{l, l_1, l_2}}{\left(l_1 + 1\right)\left(l_2 + 1\right)}
    + 2g_{l, l_1, l_2} \right]
\\ Z_{l,l_{1},l_{2}}^{m,m_{1},m_{2}} & = 
(-1)^{m}G_{l,l_{1},l_{2}}^{-m,m_{1},m_{2}}
\left[ - l + 5 
+ 2\left(l + 1\right)
\frac{\epsilon_{l, l_1, l_2}}{\left(l_1 + 1\right)\left(l_2 + 1\right)}
+ g_{l, l_1, l_2} - 2g_{l, l_2, l_1}\right]
\\ D_{l,l_{1},l_{2}}^{m,m_{1},m_{2}} & = 
(-1)^{m}G_{l,l_{1},l_{2}}^{-m,m_{1},m_{2}}
\left[
    \frac{l + 1}{2} + 
    \frac{l + 1}{2}
    \frac{\epsilon_{l, l_1, l_2}}{\left(l_1 + 1\right)\left(l_2 + 1\right)}
    -g_{l,l_{1},l_{2}}
\right]
.
\end{aligned}
\end{gather}

%% file: paperAppendixNumericalSchemes.tex
\section{Numerical schemes}
Our numerical implementation is in C++ and
based on the armadillo library 
\cite{sanderson_user-friendly_2018, sanderson_armadillo_2016}.
In addition, we use the libraries SuperLU for sparse matrix
operations \cite{SuperLU} and \cite{wigner-cpp}
for the calculation of Gaunt coefficients.

\subsection{Scheme}
\label{appendix: schemes}
We try to solve an equation of the form

\begin{equation}
    \ddot{a}_{lm}+3\frac{\dot{R}}{R}\dot{a}_{lm}-Aa_{lm}=r_{lm}
    .
\end{equation}

Here $r$ represents all nonlinear terms in the equation,
which are the most difficult to assess. 
For the purpose of the scheme, we shall treat them as some black box input.
For the time being, we make the same approximation regarding the $R, \dot R, \ddot R$ terms as the perturbations
oscillate much more rapidly. 
We present the general, half implicit scheme

\begin{gather}
    \begin{aligned}
    \frac{\dot{a}^{n+1}-\dot{a}^{n}}{dt} & +
    3\frac{\dot{R}}{R}\left(\alpha\dot{a}^{n}+\left(1-\alpha\right)\dot{a}^{n+1}\right)
    -A\left(\alpha a^{n}+\left(1-\alpha\right)a^{n+1}\right) = r \\
    \frac{a^{n+1} - {a}^{n}}{dt} &
    - \left(\alpha\dot{a}^{n} +\left(1-\alpha\right)\dot{a}^{n+1}\right) = 0
    .
    \end{aligned}
\end{gather}

Reordering terms we get

\begin{gather}
\dot{a}^{n+1}
=\frac{\dot{a}^{n}\left[1-\alpha dt\left(3\frac{\dot{R}}{R}
-dtA\left(1-\alpha\right)\right)\right]
+dtr+dtAa^{n}}{\left[1+\left(1-\alpha\right)dt
\left(3\frac{\dot{R}}{R}-dtA\left(1-\alpha\right)\right)\right]}
.
\end{gather}

We would like our scheme to be energy conserving for the pure oscillator case, 
i.e., where $\dot R = r = 0$,
and $A = -\omega_0^2$ for some real $\omega_0$. 
For a pure harmonic oscillator, the energy follows
$E \sim \dot a ^2 + \omega_0^2 a^2$. Now

\begin{gather}
    \label{eq: energy change harmonic oscillator}
    \begin{aligned}
        \Delta E & = 
        \left(\dot a_{n + 1} ^2 + \omega_0^2 a_{n + 1} ^2\right)
        - \left(\dot a_{n} ^2 + \omega_0^2 a_{n} ^2\right) \\
        & = \frac{2\alpha-1}{1+\left(1-\alpha\right)^{2}dt^{2}\omega_{0}^{2}}\omega_{0}^{2}dt^{2}E_{n}
        .
    \end{aligned}
\end{gather}

Choosing $\alpha = 1/2$ yields energy conservation.
Note that we add arbitrary dissipation and pumping terms, 
necessary for the emergence of Kolmogorov spectra.
Therefore, conservation is not of utmost importance. 
As for the radius, we evaluate radius acceleration twice,
before and after the perturbation step,
and use the average value.

\subsection{Efficiency and approximations}
\label{subsection: interaction approximation schemes}

Numerically, the most demanding task, 
by several orders of magnitude, is the evaluation of the interaction terms,
as it requires many large matrix multiplications. 
Interaction matrices are
sparse since they are proportional to the Gaunt coefficients.
The use of sparse matrix multiplication algorithms
and designated libraries such as SuperLU \cite{SuperLU}
save running time. Still,
simulations typically last several days.
We therefore approximate interaction terms, denoted $r$,
instead of directly calculating them at every step.
First, we extrapolate $r$ for short periods of time.
Suppose, $\vec r_1, ... \vec r_n$ correspond to $n$ accurate calculations of the interaction terms,
taken at times $t_1 ,... t_n$.
We approximate $r(t)$, polynomially, via

\begin{equation}
    \vec r\left(t\right) = \sum_k \vec r_k \prod_{j \neq k} \frac{t - t_j}{t_k - t_j}
    .
\end{equation}

The result is a polynomial of order $n-1$ in $t$.
To avoid overfitting, we set linear extrapolation ($n=2$). 
We measure once in $25$ steps, 
accelerating calculation by order of magnitude with no seen impact on accuracy.

Another possible approximation is the removal of the $Z$ matrix. 
For capillary waves at $l \gg 1$,
it is evident that $Z$ tensor terms are much smaller than $C$ or $D$, bringing into account that 
$\dot a_l \sim \sqrt{\alpha l^3 / R^3} a$. 
Ignoring $Z$ terms reduces running time by a factor of $1 / 3$.

At last, having reduced many matrix multiplications, 
sparse matrix additions start to take up much of the time. 
At every evaluation, we add up 
$\frac{\ddot R}{R^2} K + \frac{\alpha}{R^4} C + \frac{\dot R ^2}{R^3} X$.
For stable bubbles, $\alpha / R^4$ should remain approximately constant,
and $\dot R ^2 / R^3$ should be very small.
On the other hand, $\frac{\ddot R}{R^2}$ oscillates
with a non-negligible amplitude.
We ignore $X$ terms, much like $Z$ terms.
Furthermore, we update the matrix multiplying $a-a$ terms 
when the coefficient of either $K$ or $C$
changes by a factor of $10 \%$. 
These steps reduce approximately $2 / 3$ of sparse matrix additions.

%% file: paperAppendixTestProblems.tex
\section{Test problems}

Our code has grown rather complicated.
Therefore, we should require that it passes some tests.
In this chapter, we present two possible tests for our code.
Sadly, we have no tests checking nonlinear phenomena,
except for the predictions we set out to verify
in the first place.

\subsection{Perturbation growth in Hunter's problem}

Hunter's problem is that of a converging bubble near collapse
studied extensively in \cite{hunter_collapse_1960}.
Near collapse, we expect velocity terms
to be most dominant.
Hence, the Rayleigh-Plesset equation reduces to

\begin{equation}
    R \ddot R + \frac{3}{2}\dot R ^2 = 0
    .
\end{equation}

This equation is scaleless and exactly solvable,
with the self-similar solution

\begin{equation}
    R\left(t\right) = {\left(1 - \frac{t}{t_c}\right)}^{\frac{2}{5}}
    .
\end{equation}

Here $t_c$ is the time of the collapse. 
We now try to estimate perturbation growth.

\subsubsection{Adiabatic invariants prediction}

Let $\eta$ denote some perturbation amplitude, 
whose wavelength is much shorter than the bubble's radius.
Assuming it oscillates much faster than the bubble's
mean radius, the latter can be considered as a quasistatic
changing variable, and we can use adiabatic invariants,
such as $E / \omega$.
The energy of the perturbation must be proportional to either
$\eta^2, \eta \dot \eta, \dot \eta ^2$,
to produce leading first-order dynamics.
As the system is scaleless, 
all quadratic terms are equivalent, and we should expect
$\dot \eta \sim \frac{\dot R}{R} \eta$. Therefore,
energy and frequency must scale as

\begin{equation}
    E \sim \rho_l \dot R ^2 R  a^2
\end{equation}

and

\begin{equation}
    \omega \sim \frac{\dot R}{R}
    .
\end{equation}

Therefore,

\begin{equation}
    \frac{E}{\omega} \sim \dot R R^2 a^2 \sim const
    .
\end{equation}

Recall that $R \sim \left(-t\right)^{2 / 5}$. 
Therefore $\dot R \sim -\left(-t\right)^{-3 / 5} \sim -R^{-3/2}$, and

\begin{equation}
    a \sim R^{-1/4}
    .
\end{equation}

This result shows that perturbations are unstable near collapse.
Recall that this is a perturbative theory of first order.
Its instability demonstrates, once again, 
the importance of high-order terms for stability analysis.

\subsubsection{Analytical solutions}

Plugging Hunter's self-similar solution in 
Plesset's linear perturbation equation of motion, one finds

\begin{equation}
    \ddot a_{l,m} + 3 \frac{\dot R}{R} \dot a_{l,m} 
    + \frac{3}{2} \left(l - 1\right)\frac{\dot R ^2}{R ^2} a_{l,m} = 0
    .
\end{equation}

As there are no timescales in the problem, 
we propose a self-similar solution of the form $a \sim R^\lambda $,
reducing our differential equation to an algebraic one,

\begin{equation}
    \lambda ^2 + \frac{1}{2} \lambda + \frac{3}{2} (l-1) = 0
    .
\end{equation}

The solutions are given by

\begin{equation}
  \lambda =
    \begin{cases}
      0, -\frac{1}{2} & \text{$ l =1 $}\\
      -\frac{1}{4} \pm i \frac{1}{4} \sqrt{24 l - 25} & \text{otherwise}\\
    \end{cases}
    .       
\end{equation}

This result matches the one obtained via adiabatic invariants,
and the one obtained via a WKB approximation \cite{plesset_stability_1956}. 
For all $l > 1$, we have 
$\operatorname{Re} {\lambda_l} = -\frac{1}{4}$.
Note that adiabatic invariants hold where 
$\operatorname{Im} {\lambda_l} \gg \dot R / R$, 
or equivalently $l \gg 1$.
However, our solution indicates that the adiabatic prediction holds as early as $l = 2$.
Denote initial conditions by lower index $0$
and $\kappa_l \equiv  \frac{1}{4} \sqrt{24 l - 25}$.
For all $l > 1$ we find

\begin{equation}
    \label{eq: hunter perturbation solution}
    a = \left( \frac{R}{R_0} \right) ^ {-1/4} 
    \left[a_0 \cos{\left(\kappa_l \ln{\frac{R}{R_0}} \right)} 
    + \frac{1}{\kappa_l} \left( \frac{R_0}{\dot R_0} \dot a_0 + \frac{1}{4} a_0 \right) 
    \sin{\left(\kappa_l \ln{\frac{R}{R_0}} \right)} \right]
    .
\end{equation}
 
\begin{figure*}[!b]
    \begin{minipage}{0.45\linewidth}
        \begin{subfigure}{\textwidth}
            \includegraphics[width=\linewidth]{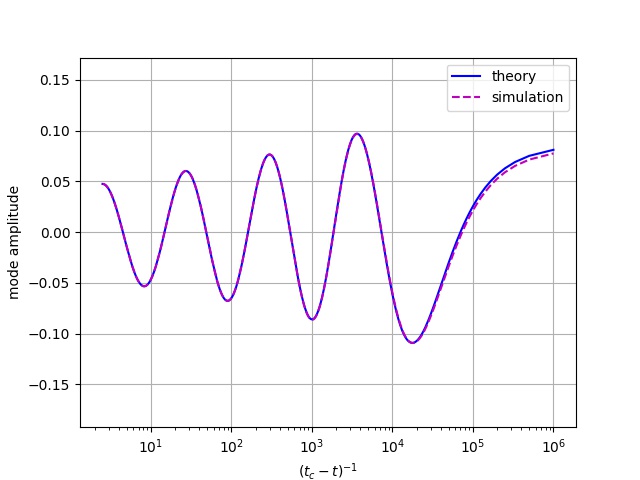}
            \caption{}\label{fig: hunter lin}
        \end{subfigure}
    \end{minipage}
    \begin{minipage}{0.45\linewidth}
        \begin{subfigure}{\textwidth}
            \includegraphics[width=\linewidth]{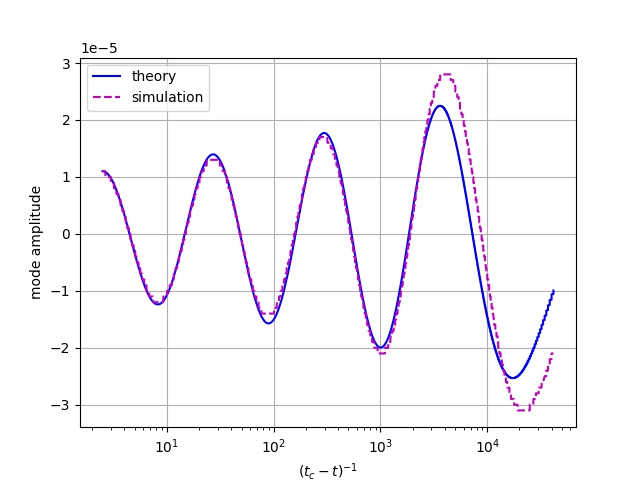}
            \caption{}\label{fig: hunter nonlin}
        \end{subfigure}
    \end{minipage}
  
    \caption{
        The figures above present simulation results
        for the amplitude of the mode $l = 30, m = 0$
        alongside the theoretical prediction by
        \eqref{eq: hunter perturbation solution}.
        Simulated with $l_{max} = 40$, $\beta = 1.0$,
        $\alpha = P_0 = P_{\infty} = 0$, 
        and no dissipation or pumping terms.
        Amplitude scale $\epsilon$ is $10 ^{-4}$ ($10 ^{-3}$)
        for a (b). $y$-axis units are arbitrary for both figures.
        }
        \label{fig:hunter_check}
\end{figure*}

We may write this equation in terms of a new, 
"natual" time of the problem, 
$T \equiv -ln{\frac{R}{R_0}}$.
$a\left(T\right)$ is of the form of 
an amplified harmonic oscillator.
As $0 < \frac{R}{R_0} \le 1$, 
$ 0 \ge T < \infty$, 
and infinitely many oscillations occur near collapse.
This analytical solution can also act as a test for our simulation.
Our simulation matches the analytical prediction
as long as perturbations are small enough.
Such is the case in figure \ref{fig: hunter lin}.
For larger perturbations,
amplitudes deviate from the linear analytic solution near collapse
due to nonlinear terms, 
as can be seen in \ref{fig: hunter nonlin}.
These graphs also verify our interpretation and 
the definition of the natural time of the problem, as 
$\left(t_c - t\right)^{-1} \propto T$.

\subsection{Collapse time}
The calculation of spherically
symmetric bubble collapse time is feasible
in several cases.
Small perturbations should not change it significantly.
Comparing simulated and theoretical
collapse times test our solution of the Rayleigh-Plesset equation.
Consider, for example, a collapsing empty spherically symmetric bubble, 
with some surface tension coefficient $\alpha$,
and $P_0 = P_{\infty} = 0$.
Moving to the unitless variables
$R = R_0 \xi$, $t = \frac{R_0}{|\dot R_0|} \chi$ and the unitless parameter 
$\theta = \frac{2 \alpha}{|\dot R_0|^2 R_0}$, the Rayleigh-Plesset equation takes the form

\begin{equation}
      \frac{\ddot \xi}{\xi} + \frac{3}{2} \frac{\dot \xi ^2}{\xi ^2} +\frac{\theta}{\xi^3} = 0
      .
\end{equation}

We make another change of variables of the form $g = \xi^\beta$, 
in an attempt to erase the nonlinear second term, 
in which we have a squared derivative. 
Such an attempt is possible when choosing $\beta = \frac{5}{2}$, 
resulting in

\begin{equation}
      \ddot g + \frac{5}{2} \theta g^{-1/5}= 0
      .
\end{equation}

Allowing us to reduce the second-order equation to a first-order equation, 
and get a differential form for $d\chi$, the unitless time. 
We finally get

\begin{equation}
      \chi_{collapse} = \frac{2}{5} \int ^1 _0 \frac{dg}{\sqrt{1 + \theta (1 - g^{4/5})}}
      .
\end{equation}

Note that in the absence of surface tension, i.e. $\theta = 0$, 
We get $\chi_{collapse} = \frac{2}{5}$, 
or, equivalently, the known $t_{collapse} = \frac{2R_0}{5|\dot R_0|}$
\cite{hunter_collapse_1960}.
This integral has an analytic solution.
Denote by $\mathcal{F}$ the incomplete elliptical integral of the first kind.
We find

\begin{equation}
    \chi_{collapse} = \frac{2}{3} \frac{1}{\theta} 
    \left[{\left(1 + \theta\right)}^{3/4}
    \theta^{-1/4} 
    \mathcal{F}\left(
    arcsin\left(\left(
        \frac{\theta}{1+\theta}\right)^{1/4}\right)
        | -1\right)
     - 1\right]
    .
\end{equation}

Computationally speaking, it is easier to evaluate
the integral form of $\chi_{collapse}$ rather than the analytic one.
We present it as it helps us reach a lower bound on the collapse time,
which is, to our knowledge, not yet known in the literature.
Our system has only positive feedback loops.
Namely, increasing bubble velocity increases its acceleration,
in a repetitive process.
Therefore, keeping $R_0$ and $\dot R_0$ constant and enlarging 
$\alpha$ must hasten collapse.
We should thus expect $\frac{\partial}{\partial\theta} \chi_{collapse} \leq 0$. 
A direct calculation yields

\begin{equation}
    \frac{\partial}{\partial\theta} \chi_{collapse} 
    = \frac{1}{2\theta (1 +\theta)} 
    + \chi \left[
        -\frac{5}{4} \frac{1}{\theta} 
        + \frac{3}{4} \frac{1}{1+\theta}\right]
    .
\end{equation}

And thus we conclude the following bound, 
tight at $\theta = 0$,

\begin{equation}
    \chi_{collapse} \geq \frac{2}{2 \theta + 5}
    .
\end{equation}

All collapse times are in $99\%$ agreement with the
calculated collapse time.
Note that this test tests only order 0 phenomena,
and is more indicative of our solution of the Rayleigh-Plesset equation,
rather than the perturbation equation.

%% file: paperAppendixInitialConditions.tex
\section{Initial conditions}

\subsection{Physical limits on initial conditions}
\label{subsection: initial conditions limit}

The series appearing in
the equations of motion we've arrived at
\eqref{eq: radius eom} and \eqref{eq: perturbation eom}
do not necessarily converge. 
Note that all coefficients are of some polynomial or rational dependence on $l$'s and generally rise (in absolute value) with them.
Series convergence constrains possible states for which our perturbative approximation applies.

To resolve divergence, one might set a series cutoff.
Either sharply, namely, demanding that all perturbations are zero from a certain $l$,
or smoothly, requiring exponential decay in amplitudes.
Either way, a cutoff defines a new scale in the system.
The other possibility is power-law scaling for perturbation amplitudes,
whose decay is fast enough for convergence. 
We choose the latter as we want our system to be as scalable as possible.
Note that setting a cutoff is inevitable. 
Firstly, as computers are not able to represent infinitely many perturbations,
setting a sharp cutoff at some $l_{max}$.
Secondly, high mode dissipation is crucial for the emergence of Kolmogorov turbulent spectra.
Kolmogorov's theory defines two scales, the pumping scale, and the viscosity scale,
the latter acting as a cutoff.
However, besides these two scales, 
there's no reason to expect any other, intermediate, length scale in the system,
relevant for behavior at the inertial interval.
As cutoff requires the definition of such a scale,
it is more reasonable to expect scaleless, power-law behavior.

We now turn to the requirements imposed on such power laws.
These are most easily arrived at when making sure the expression for $\ddot R$ converges.
Note we also ensure the convergence of interaction tensors series
along the way.

For capillary waves, $\dot a_l \sim \omega_l a_l$, Our fastest growing coefficients have a 
$O\left(l^3\right)$ divergence rate.
However, the leading order term cancels, namely 
${|\dot a|}^2 - l^3 {|a|}^2 \sim l^2 {|a|}^2$.
Note that for each $l$ there are $2l + 1 = O\left(l\right)$ modes.
Decay must therefore be at least $|a|^2 \sim l^{-4}$. 
Alternatively, $a \sim  l^{-\beta}$ For some $\beta > 2$. 
In the absence of surface tension, the fastest growing coefficient scales as
$O\left(l\right)$, similar arguments lead to the requirement that
$\beta > 1$. 
Note that the theoretical predictions for the spectrum of both capillary
and gravity waves satisfy these demands. 
For capillary waves $2\beta = 4.75$, 
and for gravity waves $2\beta = 3.5$ \cite{zakharov_kolmogorov_2012}.

\subsection{Linear waves initial conditions}
We suggest initial conditions based on the above physical requirements
regarding series convergence, and expectations from linear dynamics.
At $t = 0$ we set $R = 1$ and $\dot R = 0$.
Simple oscillatory solutions are of the form

\begin{gather}
\begin{aligned}
    a_{l,m} & 
    = a_{l,m}^{( +)} e^{i\omega_l t} 
    + a_{l,m}^{( -)} e^{ -i\omega_l t} \\
    \dot a_{l,m} & 
    = i\omega_l \left( a_{l,m}^{( +)} e^{i\omega_l t} 
    - a_{l,m}^{( -)} e^{ -i\omega_l t} \right)
    .
\end{aligned}
\end{gather}

For the local radius to be real, we simply require that
$a_{l,m}^{( -)} = \left(-1\right)^m \left[a_{l,m}^{(+)}\right] ^*$,
satisfying \eqref{eq: realness condition}.
We set $a^{(+)}$ by

\begin{equation}
    a_{l,m}^{(+)} \sim R_0 \epsilon 
    \left(\sqrt{l \left(l + 1 \right)}\right)^{ -\beta}
    \mathcal{N} \left(0, 1\right)
    .
\end{equation}

Where $\mathcal{N}$ is the standard normal distribution,
$\epsilon$ some unitless constant of order unity setting perturbation size,
and $\beta$ the scaling constant mentioned earlier.
We also put a bound of $3-4$ standard deviations,
so initial amplitudes are not too large. 
We scale with $\sqrt{l \left(l + 1 \right)}$ instead of $l$ as it is less arbitrary.
Recall that $l$-s number the eigenstates of the Laplacian, 
however, the eigenvalues themselves are $-l\left(l+1\right)$.
The choice of $l$ is somewhat arbitrary.
One could, for example, use $l` = l + 1$ to count spherical harmonics,
with Laplacian eigenvalues given by $-l` \left(l` - 1\right)$.
Hence, there is no reason to prefer $a \sim l^{-\beta}$ 
over $a \sim \left(l + 1\right)^{-\beta}$.
Eigenvalues, however, carry a physical meaning
and are independent of representation.
Therefore, we scale according to them.
For small wavelengths
$\sqrt{l \left(l + 1 \right)} \sim l \sim l'$ either way, 
and choice of scaling is not expected to be crucial.

%% file: paperAppendixDispAndPump.tex
\section{Artificial dissipation and pumping}
\label{section: artificial}
Dissipation and pumping terms are necessary for 
the creation of Kolmogorov-Zakharov spectra \cite{zakharov_kolmogorov_2012}.
We present the artificial terms we used,
taken from previously published works with a change of parameters
\cite{pushkarev_turbulence_1996, dyachenko_weak_2004}.
It is customary to add artificial terms to the equation
of motion for the canonical momenta.
Since we work in Plesset's notation of one, second-order equation, 
instead of Zakharov's two coupled first-order 
equations\footnote
{
    Here, order refers to the highest order of differentiation with respect to time.
},
we don't explicitly define the canonical momenta. 
Therefore, we exert dissipation and pumping on velocities.
We do so since Plesset's notation requires fewer interaction tensors, 
whose multiplication makes up most of the running time. 
We expect no significant difference.
Artificial coefficients are arbitrary by their very nature. 
Furthermore, for fast oscillating modes, the difference between velocities
and momenta is negligible.
As for radius dissipation, we add it to ensure stability and continuity
of dissipation at low $l$. 
We also require pure $k = \sqrt{l \left(l + 1 \right)}$ dependence of these coefficients, 
as $m$ is dependent on the arbitrary choice of the $z$ axis.
We use an operator split scheme for all artificial terms.

Dissipation terms are linear and may be divided into low mode dissipation, and high mode dissipation,
namely $\gamma_{l} = \gamma_{l} ^{low} + \gamma_{l} ^{high}$ for modes and $\gamma_{0}$, for the radius.
The dissipation split step can be solved analytically, via

\begin{gather}
    \begin{aligned}
        \dot a_{l,m}^{n + 1} & = \dot a_{l,m}^{n} e^{ - dt \gamma_{l}} \\
        \dot R^{n + 1} & = \dot R^{n} e^{ - dt \gamma_{0}}
        .
    \end{aligned}
\end{gather}

The coefficients are given by

\begin{gather}
\begin{aligned}
    \gamma_{l}^{high} & = \gamma_1 \Theta \left(k - k_d\right) \left(k^2 + k\right) \\
    \gamma_{l}^{low} & = \gamma_0 \Theta \left(k_b - k\right) 
    \left(k / k_{b} - 1\right)^ 2
    ,
\end{aligned}
\end{gather}
 
where $k_b$ and $k_d$ are derived from $l_b = 10$ and $l_d = 50$, respectively. 
$\Theta$ is the heaviside function. We choose $\gamma_1 = 10^{-1}$, $\gamma_0 = 10^{3}$. 
As for the pumping coefficients, we take the ones used by Zakharov in the original
numerical proof of the spectrum \cite{pushkarev_turbulence_1996}.
Our scheme is explicit,

\begin{gather}
    \dot a_{l,m} ^{n + 1} = \dot a_{l,m} ^{n} + dt f_{l} e^{i \Omega^n _l t}
    ,
\end{gather}

where $f_{l}$ are the coefficients, $\Omega_l$ the modified oscillation frequencies, evaluated at the beginning of the step,
and $t$ the current time. 
The modified frequency is defined by

\begin{gather}
    \Omega_l^n = \left(1 + \eta^n\right) \sqrt{\max \left(0, - A_l ^n\right)}
    ,
\end{gather}

where $\eta \sim \mathcal{N} \left(0, \sigma_\eta ^2\right)$ is a noise term. 
We choose $\sigma _ \eta = 0.05$.
The term in the square root is the effective oscillation frequency, which is time-dependent
since $\ddot R$ and $R$ vary in time, hence the upper $n$ indices. The coefficients $f_l$ are given by

\begin{gather}
    f_l = \gamma_3 e^{ - {\left(k - k_1\right)}^4 / k_2}
    ,
\end{gather}

with $\gamma_3 = 10^{-4}$ and $k_1, k_2$ are derived from $l_1 = 15$ and $l_2 = 6$, respectively. 
Note that pumping is concentrated around $k_1$ with $k_2$ setting the effective width.